# Magnetic Skyrmion Field-Effect Transistors

Ik-Sun Hong[1] and Kyung-Jin Lee[1,2,*]

[1]*KU-KIST Graduate School of Converging Science and Technology, Korea University, Seoul 02841, Korea*

[2] *Department of Materials Science and Engineering, Korea University, Seoul 02841, Korea*

**Abstract**

Magnetic skyrmions are of considerable interest for low-power memory and logic devices because of high speed at low current and high stability due to topological protection. We propose a skyrmion field-effect transistor based on a gate-controlled Dzyaloshinskii-Moriya interaction. A key working principle of the proposed skyrmion field-effect transistor is a large transverse motion of skyrmion, caused by an effective equilibrium damping-like spin-orbit torque due to spatially inhomogeneous Dzyaloshinskii-Moriya interaction. This large transverse motion can be categorized as the skyrmion Hall effect, but has been unrecognized previously. The propose device is capable of multi-bit operation and Boolean functions, and thus is expected to serve as a low-power logic device based on the magnetic solitons.

\* Corresponding email: kj_lee@korea.ac.kr (K.-J.L.)

Magnetic skyrmions are whirling spin textures, originally found in chiral magnets [1,2]. In two-dimensional systems such as thin film structures in the *x-y* plane, the magnetic skyrmion has an integer topological charge $Q$, defined by $Q \equiv \int dxdy\, \boldsymbol{m} \cdot (\partial_x \boldsymbol{m} \times \partial_y \boldsymbol{m})/4\pi$, where $\boldsymbol{m}$ is the spin order parameter with unit length. This nonzero $Q$ makes magnetic skyrmions distinct from topologically trivial spin states with $Q = 0$. Because of the topological effect, the magnetic skyrmion is expected to exhibit a high stability and to move fast at low current [3], offering intriguing possibilities for the magnetic skyrmion to be used as information carrier in memory, logic, and oscillator devices [4-12].

For device applications of magnetic skyrmions, it is important to understand the skyrmion Hall effect [13,14]: a bending of skyrmion motion with respect to the direction of the driving force (i.e., current direction). When the skyrmion Hall effect is present, the magnetic skyrmion can eventually interact with an edge of the device, resulting in a skyrmion annihilation and thus information loss. In this respect, much efforts have been devoted in eliminating the skyrmion Hall effect. The skyrmion Hall effect has the intrinsic contribution, which is independent of defects or disorders, as well as the extrinsic defect-dependent contribution. The skyrmion Hall effect is described by an effective Lorentz or Magnus force [15-17],

$$\boldsymbol{F} = Q\boldsymbol{v} \times \boldsymbol{B}, \tag{1}$$

where $\boldsymbol{v}$ is the skyrmion velocity vector, $\boldsymbol{B} = -4\pi s_{\text{net}}\hat{\boldsymbol{z}}$ is the fictitious magnetic field, and $s_{\text{net}}$ is the net spin density. The net spin density $s_{net}$ is related to the magnetic moment $M$ though $s_{\text{net}} = -M/\gamma$ where $\gamma$ is the gyromagnetic ratio. The Magnus force was predicted to become zero for antiferromagnets [18,19] or antiferromagnetically coupled ferrimagnets [20] where $s_{\text{net}}$ is zero. This prediction was recently confirmed by an experiment using antiferromagnetically coupled ferrimagnets where $s_{\text{net}}$ vanishes at a special condition called the angular momentum compensation point [21].

In this work, we take a different view on the skyrmion Hall effect for applications: we actively exploit the skyrmion Hall effect, instead of vanishing it. For this purpose, we focus on the extrinsic defect-dependent contribution to the skyrmion Hall effect, which is known to be nonnegligible [22-24]. To date, most theoretical and numerical studies on the extrinsic skyrmion Hall effect have focused on point defects that are assumed to create featureless pinning potential [22,23]. Each point defect generates a *small* transverse motion of a magnetic skyrmion. Because of random nature of point-defect distributions in real samples, such point defects result in uncontrollable local transverse motions. We here consider a planar defect of which size is much larger than the skyrmion size and its pinning potential is controlled by a gate voltage. In contrast to a point defect, this planar defect causes a *large* transverse motion of a magnetic skyrmion in a controllable manner, providing the key working principle of magnetic skyrmion field-effect transistor, which we propose in this work.

To be applicable for field-effect transistors where the gate-controlled operation is essential, we consider a pinning potential caused by spatially inhomogeneous Dzyaloshinskii-Moriya interaction (DMI). Since magnetic skyrmions are stabilized by the DMI, the spatial inhomogeneity of DMI may have an important role in the skyrmion Hall effect. Moreover, the pinning potential is either a barrier or a well, depending on its sign, as exemplified by our recent work on the effect of spatially modulated DMI on magnetic domain wall motion [25]. We showed in Ref. [25] that the domain wall motion is strongly affected by the sign of the DMI difference (i.e., $\delta_D = D_1 - D_2$) when the domain wall moves from a region to another region, where each region has a different DMI ($D_1$ for the region 1 and $D_2$ for the region 2). Therefore, it is reasonable to expect that the skyrmion motion is also affected by the sign of the DMI difference. Moreover, recent experiments demonstrated the gate-controlled DMI [26,27]. In particular, Ref. [26] reported that the interfacial DMI can be controlled as large as 100 % by the gate voltage, offering the gate-controlled pinning potential originating from the DMI.

In this work, we numerically investigate the effect of a DMI defect on the skyrmion Hall effect. We show results not only for a planar defect of which size is much larger than the skyrmion size, but also for a point defect of which size is smaller than the skyrmion size, because both types of defect exhibit similar mechanism for the transverse motion. We find that the transverse motion of skyrmion deflected by the defect can be decomposed to *small* and *large* motions. The sign of the *small* motion depends only on the topological charge $Q$ of the skyrmion. On the other hand, the sign of the *large* motion depends on the signs of both $Q$ and $\delta_D$. At the end of this paper, we show that the latter *large* motion can be used to construct gate-controlled multibit skyrmion field-effect transistors.

We perform two-dimensional micromagnetic simulation by numerically solving the Landau-Lifshitz-Gilbert (LLG) equation including the adiabatic and non-adiabatic spin-transfer torques,

$$\frac{d\hat{\mathbf{m}}}{dt} = -\gamma \hat{\mathbf{m}} \times \mathbf{H}_{eff} + \alpha \hat{\mathbf{m}} \times \frac{d\hat{\mathbf{m}}}{dt} + b_j \frac{\partial \hat{\mathbf{m}}}{\partial x} - \beta b_j \hat{\mathbf{m}} \times \frac{\partial \hat{\mathbf{m}}}{\partial x}, \qquad (2)$$

where $\hat{\mathbf{m}}$ is the unit vector along the magnetization, $\mathbf{H}_{eff}$ is the effective field including the exchange, anisotropy, DMI, and dipolar fields, $\alpha$ is the damping constant, $b_j = (P\mu_B J/eM_S)$ is the spin-current velocity, $P$ is the spin polarization, $\mu_B$ is the Bohr magneton, $J$ is the current density, $e$ is the electron charge, $M_S$ is the saturation magnetization, and $\beta$ is the nonadiabaticity. We assume that the electrical current flows along the *x*-axis, and the thickness direction is in the *z*-axis. We use following parameters: the exchange stiffness constant $A =$ $1.3 \times 10^{-6}$ erg/cm, the uniaxial anisotropy constant along the $z$ easy axis $K = 7 \times 10^6$ erg/cm$^3$, the DMI constant $D = 2.5$ erg/cm$^2$, the saturation magnetization $M_S = 580$ emu/cm$^3$, the spin polarization $P = 0.7$, the damping constant $\alpha = 0.1$, the current density $J = 4 \times 10^7$ A/cm$^2$, $\gamma = 1.76 \times 10^7$ $Oe^{-1}s^{-1}$, and $|s_{net}| = 3.295 \times 10^{-5}$ $erg \cdot s/cm^3$. We assume $\alpha = \beta$ in order to focus on the extrinsic skyrmion Hall effect originating from the interaction with defects. For the condition of $\alpha = \beta$, the direction of current-driven skyrmion motion is parallel to the current direction ($x$-axis) in the absence of defects. We characterize a point DMI defect as $D(r) = D - \delta_D \exp(-(r - r_0)^2/\Delta)$, where $D$ is the homogeneous DMI except for the point defect, $\delta_D$ is the DMI difference between the homogeneous region and the defect, $r_0$ is the center position of point defect, and $\Delta$ is a parameter proportional to the defect area. We characterize a planar defect in the $xy$ plane as $D(x) = D - \delta_D \Theta(x)\Theta(X_0 - x)$, where $\Theta(x)$ is the Heaviside step function and $X_0$ is the width of a planar defect along the $x$ direction.

We first show simulation results for a point defect. Figure 1 shows trajectories of spin-transfer-torque-driven skyrmion motion depending on the initial location of a magnetic skyrmion. The initial locations are on the left side of a point defect and are varied along the vertical direction (= $y$ direction). A circle at the center corresponds to a point defect of which sign of $\delta_D$ is indicated by color (i.e., red means positive whereas blue means negative). We consider a point defect of which size ($\approx$ 10 nm) is smaller than the skyrmion diameter ($\approx$ 20 nm). In Fig. 1(a) and (c), the skyrmion has a positive topological charge ($Q > 0$) whereas in Fig. 1(b) and (d), it has a negative topological charge ($Q < 0$) by reversing $\boldsymbol{m}$ to $-\boldsymbol{m}$.

When a skyrmion moves from left to right, it is scattered by the point defect. When $Q > 0$ and $\delta_D > 0$ [Fig. 1(a)], for instance, the skyrmion trajectory is largely deflected by the point defect in the $-y$ direction on the left side of the defect, while it is largely deflected in the $+y$ direction on the right side of the defect. We call this large transverse deflection near the point defect as the *large* motion. After this *large* motion disappears, the skyrmion is slightly shifted in the $-y$ direction for Fig. 1(a) (i.e., when comparing the initial location on far left to the final location on far right of the point defect), which we call the *small* motion. One can find general dependences of the *large* and *small* motions on the signs of $Q$ and $\delta_D$ as summarized in the following: the sign of *large* motion is determined by the sign of $Q \cdot \delta_D$ whereas the sign of *small* motion is determined solely by the sign of $Q$. Given that ferromagnets have fixed net spin density $s_{\text{net}}$, one finds from Eq. (1) that the sign of the effective Magnus force depends only on the topological charge $Q$. Therefore, the *small* motion, of which sign depends only on the sign of $Q$, represents the conventional extrinsic skyrmion Hall effect induced by a point defect. This interpretation based on Eq. (1) indicates that the *large* motion has an origin other than the skyrmion Hall effect. As we show below, the origin of the *large* motion is better explained for a planar defect, so that we will discuss about it below.

We next show simulation results for a planar defect. Figure 2 shows trajectories of spin-transfer-torque-driven skyrmion motion depending on $Q$ and $\delta_D$, where a planar defect is indicated by a shaded region covering from $x = 100$ nm to $x = 200$ nm. This shaded region has nonzero $\delta_D$, of which sign is indicated by color. Here we fix the initial location of a skyrmion at $(x, y) = $ (75 nm, 100 nm). We find that the transverse motion of skyrmion again consists of the *large* (= $\Delta y_1$) and *small* (= $\Delta y_2$) motions. As for a point defect (Fig. 1), the sign of *large* motion is determined by the sign of $Q \cdot \delta_D$ whereas the sign of *small* motion is determined solely by the sign of $Q$.

The *large* motion can be explained by an effective DMI field originating from the spatially inhomogeneous DMI at the interface between the homogeneous region and the planar defect [28]. The DM energy at the interface with a planar defect is given by

$$E_{DM} = D\hat{\mathbf{y}} \cdot (\hat{\mathbf{m}}_{i-1} \times \hat{\mathbf{m}}_i) + (D - \delta_D)\hat{\mathbf{y}} \cdot (\hat{\mathbf{m}}_i \times \hat{\mathbf{m}}_{i+1})$$

$$= D\hat{\mathbf{y}} \cdot (\hat{\mathbf{m}}_i \times (\hat{\mathbf{m}}_{i+1} - \hat{\mathbf{m}}_{i-1})) - \delta_D \hat{\mathbf{y}} \cdot (\hat{\mathbf{m}}_i \times \hat{\mathbf{m}}_{i+1}). \quad (3)$$

From the above equation, we obtain the effective DM field at the step as

$$\mathbf{H}_{DM,eff} = -\frac{\delta E_{DM}}{\delta M} = \frac{1}{M_S}\left[D\hat{\mathbf{y}} \times \left(\frac{\partial \hat{\mathbf{m}}}{\partial x}\right) + \frac{\delta_D}{2a}(\hat{\mathbf{y}} \times \hat{\mathbf{m}})\right], \quad (4)$$

where $a$ is the lattice constant. On the right hand side of Eq. (4), the first term is the ordinary DMI field whereas the second term is an additional DMI field originating from spatially inhomogeneous DMI, which is proportional to $\delta_D$. We note that the symmetry of the additional DMI field is similar to that of the non-equilibrium (i.e., current-driven) damping-like spin-orbit torque [28], but this additional DMI field is an equilibrium one. Because the additional DMI field is independent of the current, one can identify it by moving a skyrmion without current injection: e.g., using oscillating magnetic field gradient [29]. When the skyrmion passes through the planar defect, it experiences this effective damping-like spin-orbit torque [i.e., a torque due to the second term of Eq. (4)]. The skyrmion dynamics driven by a combined action of adiabatic, non-adiabatic spin-transfer torques, and damping-like spin-orbit torque is given by

$$\mathcal{G} \times (\mathbf{v}_s - \mathbf{v}_d) + \mathcal{D}\ (\beta \mathbf{v}_s - \alpha \mathbf{v}_d) + \mathbf{F}^{DMI} = 0, \quad (5)$$

where $\mathbf{v}_s = (v_s, 0)$ is the spin-current velocity, $\mathbf{v}_d = (v_x, v_y)$ is the skyrmion velocity, and $\mathbf{F}^{DMI} = (F_x, F_y) = (-(\delta_D/2aM_S)\pi\lambda, 0)$ is a force term induced by the inhomogeneous DMI where $\lambda$ is characteristic length of the skyrmion. The first term is the Magnus force with gyromagnetic coupling vector $\mathcal{G} = \hat{\mathbf{z}} \int \hat{\mathbf{m}} \cdot (\partial_x \hat{\mathbf{m}} \times \partial_y \hat{\mathbf{m}}) \, d^2r = \mathcal{G}\hat{z}$. The second term is the dissipative force with dissipative tensor $\mathcal{D} = \delta_{ij} \int (\partial_i \hat{\mathbf{m}} \cdot \partial_j \hat{\mathbf{m}}) \, d^2r = \delta_{ij}\mathcal{D}$. One can investigate the skyrmion dynamics driven by current-induced spin-orbit torque by setting $v_s = 0$ and adding $\boldsymbol{F}^{SOT}$ [4] in Eq. (5), for which effects of $\boldsymbol{F}^{DMI}$ and $\boldsymbol{F}^{SOT}$ are simply additive.

For $\alpha = \beta$, the skyrmion velocities obtained from Eq. (5) are given as

$$v_x = v_s - \frac{\alpha \mathcal{D}}{\mathcal{G}^2 + (\alpha \mathcal{D})^2} \left( \frac{\delta_D}{2aM_S} \pi \lambda \right), \tag{6}$$

$$v_y = \frac{\mathcal{G}}{\mathcal{G}^2 + (\alpha \mathcal{D})^2} \left( \frac{\delta_D}{2aM_S} \pi \lambda \right). \tag{7}$$

Equation (7) shows that the sign of $v_y$ is determined by the sign product of $\mathcal{G}$ (thus, $Q$) and $\delta_D$, consistent with numerical result. From Eqs. (6) and (7), moreover, one finds that the ratio $v_y/v_x$ proportional to the net skyrmion Hall angle decreases with increasing the current density (i.e., $v_s$). We note that this *large* transverse motion induced by nonzero $\delta_D$ can be categorized as the skyrmion Hall effect, but has been unrecognized so far.

Our result suggests that exploiting the *large* motion could be useful for additional functionality of skyrmion devices. The *large* motion induced by a planar defect combined with the gate-controlled DMI, which was experimentally demonstrated [26], allows us to largely tune the transverse motion of skyrmion. Here we propose a skyrmion field-effect transistor using the gate-controlled DMI, shown in Fig. 3(a). By applying a gate voltage $V_{gate}$, the DMI of the region (= region 1) beneath the gate electrode varies, while the DMI of other regions (= region 2) is not affected by the gate voltage. As a result, a DMI step is formed between the region 1 and region 2, which supplies an effective damping-like spin-orbit torque on the skyrmion [i.e., a torque due to the second term of Eq. (3)]. Since the effective damping-like spin-orbit torque originating from the inhomogeneous DMI is proportional to $\delta_D$, which is controllable by a gate voltage, the skyrmion at the boundary shows *large* motion with an amount depending on a gate voltage. As a result, a skyrmion field-effect transistor can have multi-bit states classified by the skyrmion position in the *y* axis at the region 1, which is controllable by the magnitude and sign of the gate voltage (i.e., the magnitude and sign of $\delta_D$ in region 1). The skyrmion position or, equivalently, the presence/absence of skyrmion is detected by, for instance, tunnel magnetoresistances of magnetic tunnel junctions formed at the detectors in Fig. 3(a). Figure 3(b) shows schematic diagrams of four-bit operations of a skyrmion field-effect transistor. Figure 3(c) shows that the *large* motion (= $\Delta y_1$) is about 30 nm at $\delta_D = 0.1$ erg/cm$^2$. Given the assumed homogeneous DMI $D = 2.5$ erg/cm$^2$, therefore, $\delta_D/D$ (×100 %) for inducing $\Delta y_1 = 30$ nm requires the gate-controlled DMI change of only 4%. Because of the linear proportionality between $\Delta y_1$ and $\delta_D$ [Fig. 3(c)], a larger $\Delta y_1$ is easily achievable by increasing $\delta_D$, demonstrating that the gate-controlled DMI can be efficiently employed in the proposed skyrmion field-effect transistor. Furthermore, the Boolean logic functions can be implemented in the proposed device. Figure 3(d) show a unit cell of the skyrmion field-effect transistor for this purpose. A magnetic skyrmion (black dot), initially located in the left region, can move to the right region when the DMI gate is on (i.e., V$_G$ = "1"), whereas it cannot when the DMI gate is off (i.e., V$_G$ = "0"). When connecting two unit cells in series (parallel), "AND" ("OR") Boolean function is realized. Other Boolean functions can be implemented in a similar way.

In conclusion, we propose a skyrmion field-effect transistor that utilizes the *large* transverse motion of magnetic skyrmion, induced by the gate-controlled DMI. An important merit of the proposed skyrmion field-effect transistor is the capability of multi-bit operation by simply varying the gate voltage. Because magnetic skyrmions move fast at a low current [3], our proposal of the multi-bit skyrmion field-effect transistors will be useful for low-power logic devices based on magnetic solitons.

**Acknowledgement**

This work is supported by the National Research Foundation of Korea (NRF) grant funded by the Korea government (MSIP) (2015M3D1A1070465, 2017R1A2B2006119) and KU-KIST Graduate School of Converging Science and Technology Program.

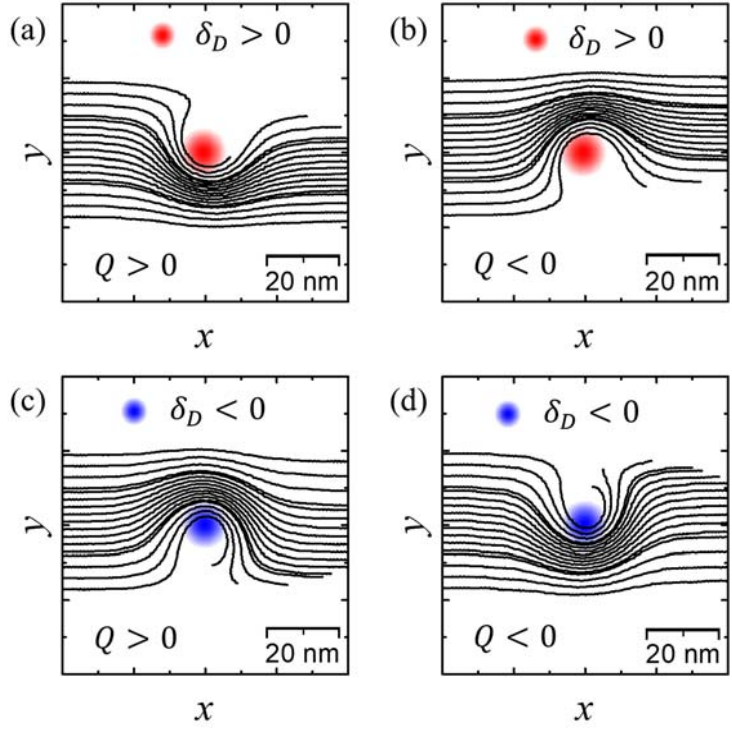

Figure 1. Transverse motion of a magnetic skyrmion induced by the interaction with a point defect. (a) $Q > 0$ and $\delta_D > 0$, (b) $Q < 0$ and $\delta_D > 0$, (c) $Q > 0$ and $\delta_D < 0$, and $Q < 0$ and $\delta_D < 0$. Red and blue circles indicate the DMI point defects. Lines indicate the skyrmion trajectories for skyrmions moving from left to right.

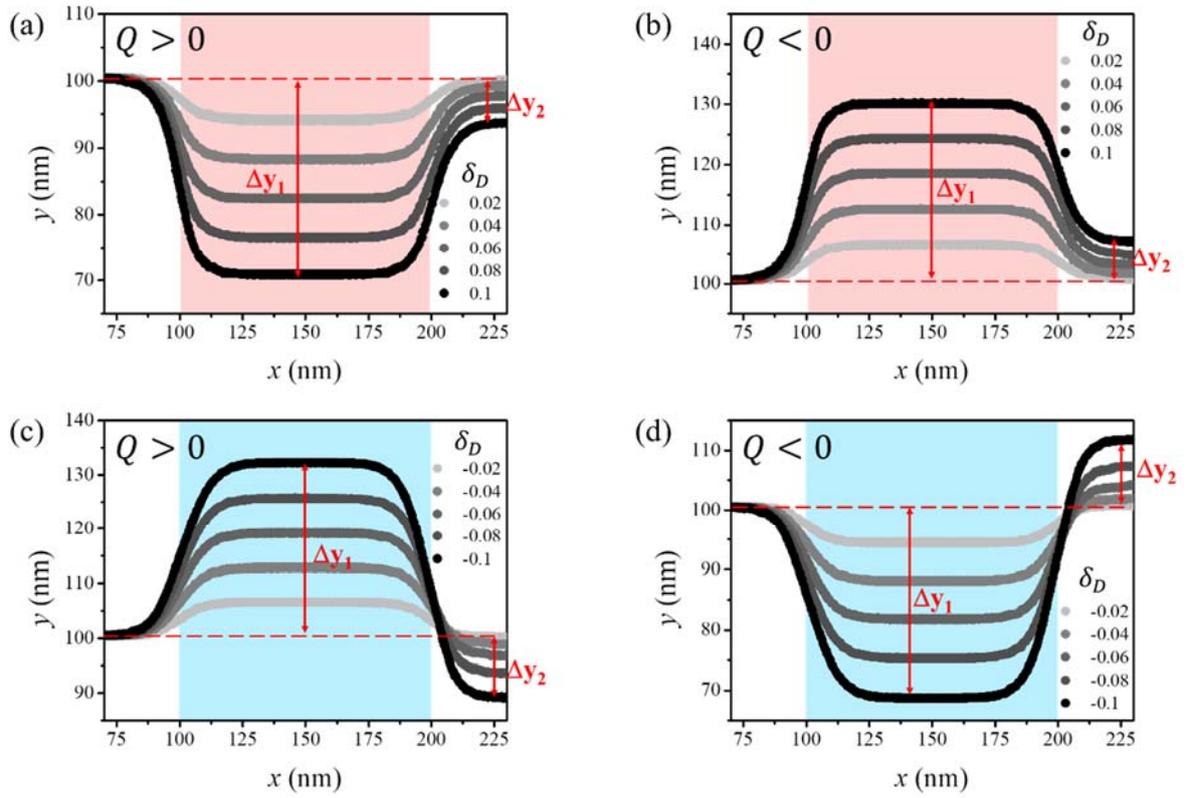

Figure 2. Transverse motion of a magnetic skyrmion induced by the interaction with a planar defect. (a) $Q > 0$ and $\delta_D > 0$, (b) $Q < 0$ and $\delta_D > 0$, (c) $Q > 0$ and $\delta_D < 0$, and $Q < 0$ and $\delta_D < 0$. Red and blue regions indicate the DMI planar defects. Lines indicate the skyrmion trajectories for a skyrmion moving from left to right. $\Delta y_1$ and $\Delta y_2$ correspond to the *large* and *small* transverse motions, respectively.

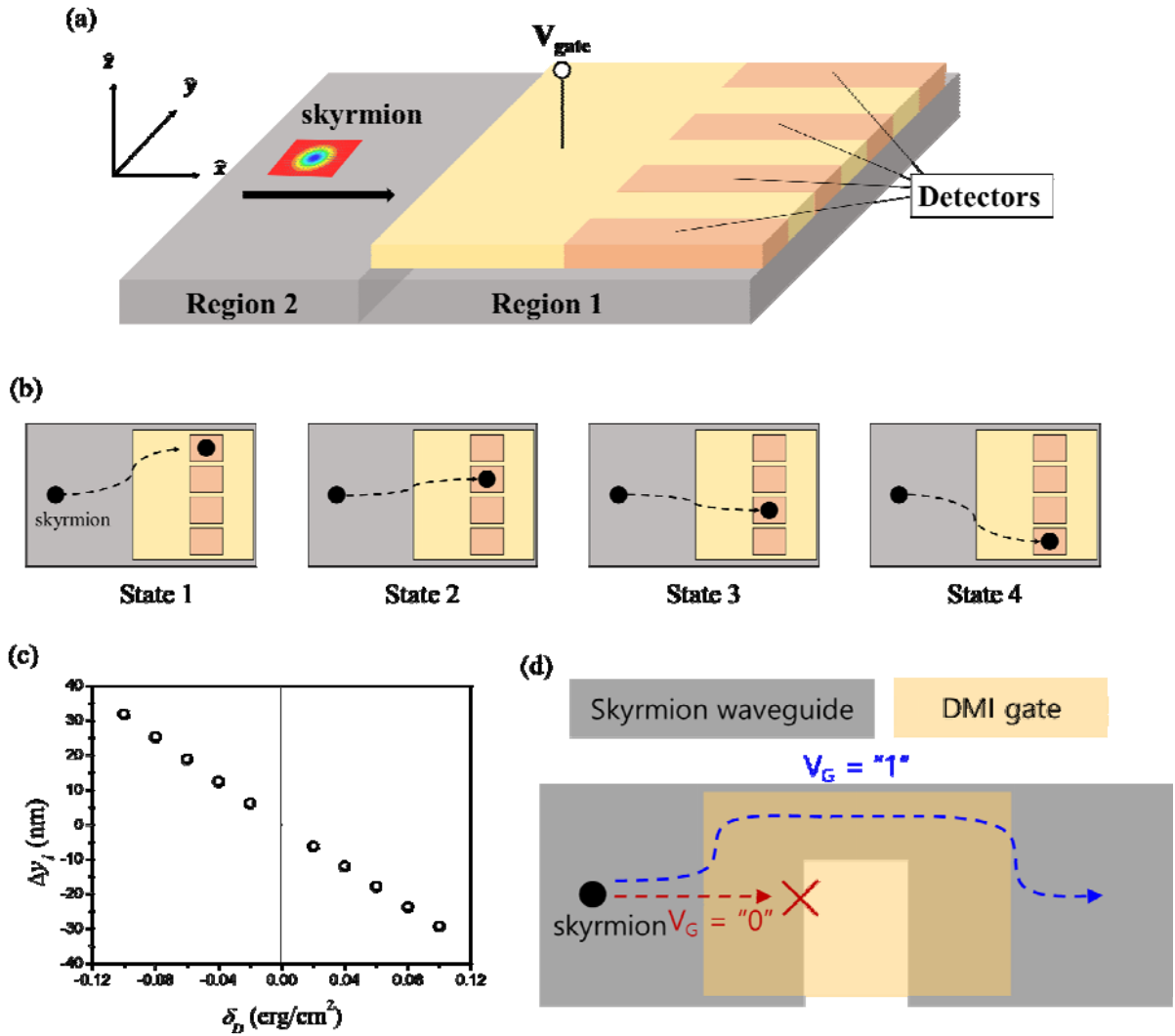

Figure 3. (a) Schematic of a skyrmion field-effect transistor. The left part of the device, which is not covered by a gate electrode, is the region 2 and the right part, which is covered by a gate electrode, is the region 1. Orange-graded regions on the right are the skyrmion detectors, which can be formed by magnetic tunnel junctions, for example. (b) Multi-bit operation of skyrmion field-effect transistor. Each of the state 1 to state 4 can be selected by applying a different gate voltage, which causes a different $\delta_D$ and consequently, different *large* motion. (c) The *large* motions ($\Delta y_1$) as a function of $\delta_D$ for a skyrmion field-effect transistor. (d) A unit cell of the skyrmion field-effect transistor for Boolean function.

16